\documentclass[12pt]{iopart}

\usepackage[vietnamese, english]{babel}
\usepackage{braket}

\usepackage{graphicx}%
\usepackage{multirow}%
\expandafter\let\csname equation*\endcsname=\relax 
\expandafter\let\csname endequation*\endcsname=\relax 
\usepackage{amsmath,amssymb,amsfonts,physics}%
\usepackage{amsthm}%
\usepackage{mathrsfs}%
\usepackage[title]{appendix}%
\usepackage{xcolor}%
\usepackage{textcomp}%
\usepackage{manyfoot}%
\usepackage{booktabs}%
\usepackage{algorithm}%
\usepackage{algorithmicx}%
\usepackage{algpseudocode}%
\usepackage{listings}%

\theoremstyle{thmstyletwo}%

\theoremstyle{thmstylethree}%

\begin{document}

\title[Few-body precursors of topological frustration]{Few-body precursors of topological frustration}

\author{Federico~Raffaele~De~Filippi}
\address{Department of Physics, University of Trieste, 34127 Trieste, Italy}
\ead{federicoraffaele.defilippi@ds.units.it}

\author{Antonio~Francesco~Mello}
\address{International School for Advanced Studies (SISSA), 34136 Trieste, Italy}
\ead{amello@sissa.it}

\author{Daniel Sacco Shaikh}
\address{Dipartimento di Fisica, Università degli Studi di Genova, via Dodecaneso 33, 16146, Genova, Italy}
\ead{4526348@studenti.unige.it}

\author{Maura Sassetti}
\address{Dipartimento di Fisica, Università degli Studi di Genova, via Dodecaneso 33, 16146, Genova, Italy}
\address{CNR spin, via Dodecaneso 33, 16146, Genova, Italy}
\ead{maura.sassetti@unige.it}

\author{Niccolò Traverso Ziani}
\address{Dipartimento di Fisica, Università degli Studi di Genova, via Dodecaneso 33, 16146, Genova, Italy}
\address{CNR spin, via Dodecaneso 33, 16146, Genova, Italy}
\ead{niccolo.traverso.ziani@unige.it}

\author{Michele~Grossi}
\address{European Organization for Nuclear Research (CERN) Geneva 1211, Switzerland}
\ead{michele.grossi@cern.ch}
\vspace{10pt}
\begin{indented}
\item[]January 2024
\end{indented}

\begin{abstract}
Quantum spin chains - the prototypical model for coupled two-level systems - offer a fertile playground both for fundamental and technological applications, ranging from the theory of thermalization to quantum computation. The effects of frustration induced by the boundary conditions have recently been addressed in this context. In this work, we analyze the effects of such frustration on a few spin system and we comment the strong even-odd effects induced in the ground state energy. The purpose of this work is to show that such signatures are visible on current quantum computer platforms.
\end{abstract}

%
%
%
%
%



\maketitle

\section{Introduction}\label{sec1}

The theoretical importance and the technological potential of coupled multilevel quantum systems are substantial. They range from fundamental issues related to thermalization \cite{Ter1,Ter2,Ter3,Ter4} and information or energy spreading \cite{QI1}, to the realization of quantum bits \cite{Thore} and quantum batteries \cite{Dario}, and towards the simulation of complex quantum systems \cite{Sim}. In this context, recently, a new intriguing phenomenon has been discovered: Topological Frustration (TF) \cite{Giampaolo_area_law, torre2023longrange}. Qualitatively speaking, TF amounts to the fact that some spin chains with $N$ spins obeying periodic boundary conditions behave very differently, even for large $N$, depending on the parity of $N$. To understand the physical content, it is useful to start with a very intuitive classical argument: imagine a $N$ site classical antiferromagnetic Ising model in the absence of applied fields, compactified on a circle. If $N$ is even, the lowest energy state is the one with neighbouring spins antiparallel to each other. The degeneracy is two: the state with all the spins flipped has the same energy. If $N$ is odd, on the other hand, it is not possible, due to the boundary conditions, to have all spins antiparallel to their neighbours. TF is hence at play. The lowest energy state is in this case the one where all spins except for one are antiparallel to their neighbours. The important fact is now that the degeneracy of the lowest energy state is now $2N$ instead of 2. At the quantum level, when for example a transverse field is switched on, the even $N$ case is characterized in the thermodynamic limit $N\to\infty$ by a gap between the (doubly degenerate) ground state and the excited states, while in the odd $N$ case the systems becomes gapless, with the gapless modes originating from the hybridization of the $2N$ states that are degenerate at the classical point, that is in the absence of applied fields \cite{Dong_2016}. The implications of this simple observation are complex and far reaching \cite{Frustration_of_being_odd, fate_of_local_order, GiampaTorre, Giampaolo_area_law, Torre_loschmidt, torre2023longrange, Catalano22, Odavic_SciPostPhys, catalano2023frustrating, TF_modify_the_nature_of_QPT}. Boundary quantum phase transitions that are absent in the even $N$ case show up for odd $N$, marking the onset of ground states with finite global magnetization in the antiferromagnetic case, or the spontaneous breaking of translational invariance \cite{QPT_induced_by_TF, articolo_tesi, shaikh2023phase}. At the non-equilibrium level, signatures of TF are also visible in the Loschmit echo \cite{Torre_loschmidt}. Although the best way to take the thermodynamic limit in systems with TF is under intense debate \cite{Catalano22, shaikh2023phase}, what is clear is that systems with finite $N$ presenting TF have a very strong dependence of their properties on the parity of $N$. This fact is what we theoretically and experimentally inspect in this work. In particular, we analyze analytically the effects of topological frustration in the XY quantum spin chain with $N=5,6$ spins. Among other effects, in particular, we show that that the ground state energy has a way stronger dependence on the applied magnetic field in the presence of TF (antiferromagnetic coupling, odd $N$), than in the non-frustrated cases, as first observed by Dong \textit{et al.} in \cite{Dong_2016, Dong_2017}. To go, for the first time, beyond the theoretical description of TF, we then implement a special case of the model discussed theoretically on an IBM quantum computer where the role of the spins is taken by trasmon qubits. Remarkably, we confirm the peculiarity of the system with TF in terms of ground state energy.\\
On one hand, our results confirm the robustness of the physics related to TF in real life contexts, where thermal fluctuations are unavoidable and exact symmetries in the coupling parameters essentially do not exist. On the other hand, they open the way to the conception of nanodevices based on TF, where by only changing $N$ by one, the properties of the system change dramatically.\\
The rest of the article is structured as follows. In Sec.2, we present and recap the solution of the model. We also discuss the features of topological frustration, resuming some the most relevant results in the literature so far. In Sec.3 we discuss the effects of topological frustration on finite size systems, focusing in particular on the $N=5,6$ cases. In Sec. 4 we present our experimental results, and finally in Sec. 5 we illustrate our conclusions.
\section{The XY model and its topological frustration}\label{Section model and diagonalization}
\subsection{Model}
The model under investigation is the so called XY quantum chain in a transverse field \cite{Franchini_2017, katsura, LIEB1961407, NIEMEIJER1967377, NIEMEIJER1968313}. The Hamiltonian $H$ is given by
\begin{align}\label{xy model}
    H = \frac{J}{2}\sum_{j=1}^N \left[\left(\frac{1+\gamma}{2}\right)\sigma_j^x\sigma_{j+1}^x + \left(\frac{1-\gamma}{2}\right)\sigma_j^y\sigma_{j+1}^y + h\sigma_j^z\right],
\end{align}
where $N$ is the number of sites, $J$ is the energy scale, $\gamma$ is a parameter quantifying the anisotropic way in which nearest neighbour spins interact, $h$ is an external magnetic field along $z$ direction and $\sigma_j^\alpha$ (with $\alpha=x,y,z$) are Pauli matrices in the usual representation. Clearly one has
\begin{equation}
    [\sigma_j^\alpha, \sigma_k^\beta] = 2i \epsilon^{\alpha\beta\gamma} \sigma_j^\gamma \delta_{jk}.
\end{equation}
Crucially for the following, we impose periodic boundary conditions
\begin{equation}\label{PBC}
    \sigma_{N+1}^\alpha \equiv \sigma_1^\alpha.
\end{equation}

We now proceed with the diagonalization of the Hamiltonian, which is known in literature. Indeed, the model defined in \eqref{xy model} is one of the most famous integrable models and was first introduced and solved in the $h=0$ case by Lieb, Schultz and Mattis in \cite{LIEB1961407} and, successively, in \cite{katsura} with a finite external field. Since however the TF regime requires some care, we find it worthy to report here the explicit solution.\\
First of all, thanks to the symmetry properties of \eqref{xy model}, we restrict our analysis, without loss of generality, only to the first quarter of the parameters space $(h,\gamma)$, \textit{i.e.} $h,\gamma \geq 0$. We then introduce the Wigner-Jordan transformation \cite{Jordan1928berDP}, in which the spin operators are non-locally mapped onto spinless fermions. One has
\begin{align}\label{def trasf Wigner-Jordan}
    \sigma_j^+ \equiv \exp\left\{i\pi \sum_{l=1}^{j-1}\psi_l^\dagger \psi_l\right\}\psi_j \qquad j=1,...,N,
\end{align}
with $\sigma_j^+=(\sigma_j^x+i\sigma_j^y)$, and the other operators obtainable from the commutation relations. Here, we have the anticommutation relations
\begin{equation}\label{ferm}
\psi_l^\dagger\psi_m+ \psi_m\psi_l^\dagger=\delta_{l,m},
\end{equation}
with $l$, $m$ running over the sites of the chain and $\delta_{i,j}$ the Kronecker symbol, while all other anticommutation relations are zero.\\
With respect to the new basis of operators, the Hamiltonian $H$ reads 
\begin{align}
    H = &\frac{J}{2}\sum_{j=1}^{N-1}(\psi_{j+1}^\dagger \psi_j + \psi_j^\dagger \psi_{j+1} + \gamma \psi_{j+1}\psi_j + \gamma \psi_j^\dagger \psi_{j+1}^\dagger) +\frac{JhN}{2} -Jh\sum_{j=1}^N \psi_j^\dagger \psi_j + \nonumber\\
    &-\frac{J}{2}\Pi^z(\psi_1^\dagger \psi_N + \psi_N^\dagger \psi_1  +\gamma \psi_1\psi_N + \gamma \psi_N^\dagger \psi_1^\dagger),\label{H xy in fne di psi}
\end{align}
with
\begin{equation}\label{parity operators}
    \Pi^\alpha \equiv \bigotimes_{l=1}^N \sigma_l^\alpha ,
\end{equation}
the parity operator. We note that the Hamiltonian when written in terms of the fermions is highly non-local and non-linear. However, since
\begin{equation}\label{z-parity simmetry}
    [H,\Pi^z] =0,
\end{equation}
one can decompose the Hamiltonian as
\begin{equation}\label{z parity constraint}
    H = \frac{1+\Pi^z}{2} H^+ \frac{1+\Pi^z}{2} + \frac{1-\Pi^z}{2} H^- \frac{1-\Pi^z}{2},
\end{equation}
where
\begin{align}
    H^\pm = &\frac{J}{2}\sum_{j=1}^{N-1}(\psi_{j+1}^\dagger \psi_j + \psi_j^\dagger \psi_{j+1} + \gamma \psi_{j+1}\psi_j + \gamma \psi_j^\dagger \psi_{j+1}^\dagger) +\frac{JhN}{2} -Jh\sum_{j=1}^N \psi_j^\dagger \psi_j + \nonumber\\
    &\mp\frac{J}{2}(\psi_1^\dagger \psi_N + \psi_N^\dagger \psi_1  +\gamma \psi_1\psi_N + \gamma \psi_N^\dagger \psi_1^\dagger)\\
    \equiv& \frac{J}{2}\sum_{j=1}^N(\psi_{j+1}^{(\pm)\dagger}\psi^{(\pm)}_j + \psi_j^{(\pm)\dagger}\psi^{(\pm)}_{j+1} +\gamma\psi^{(\pm)}_{j+1}\psi^{(\pm)}_j + \gamma\psi_j^{(\pm)\dagger}\psi_{j+1}^{(\pm)\dagger }  -2h\psi_j^{(\pm)\dagger}\psi^{(\pm)}_j)+\frac{JhN}{2}.\label{H^pm(psi^pm)}
\end{align}
In the last equality we defined
\begin{align}\label{def psi^pm}
    \begin{cases}
    \psi_j^{(\pm)} \equiv \psi_j  \\[1.6 ex]
    \psi^{(\pm)}_{j+N} \equiv \mp\psi^{(\pm)}_{j}
    \end{cases}
    \qquad j= 1,...,N.
\end{align}
We are now in the position of solving the Hamiltonian. To do so, we now switch to Fourier space, with the convention
\begin{equation}\label{def trasf fourier fermioni di Wigner-Jordan}
    \psi_j^{(\pm)} \equiv \frac{1}{\sqrt{N}} e^{i\frac{\pi}{4}}\sum_{q\in \Gamma^\pm}e^{iqj}\psi_q,
\end{equation}
where
\begin{equation}\label{def Gamma^pm}
    \Gamma^+ \equiv \left\{\frac{2\pi}{N}\left(k+\frac{1}{2}\right)\right\} \qquad \Gamma^- \equiv \left\{\frac{2\pi}{N}k\right\}\qquad k=0,...,N-1.
\end{equation}
These two sets will play a crucial role in the mathematical derivation of even-odd effects in this model.
\\In Fourier space one finds
\begin{align}
    H^{\pm} &= -J\sum_{q\in \Gamma^\pm}(h-\cos q)\psi_q^\dagger \psi_q -\frac{J\gamma}{2}\sum_{q\in\Gamma^\pm}\sin q (\psi_q \psi_{2\pi-q} + \psi^{\dagger}_{2\pi-q} \psi_q^\dagger) + \frac{JhN}{2}\nonumber\\
    &= \frac{J}{2}\sum_{q\in\Gamma^\pm}\begin{pmatrix}
    \psi_q^{\dagger} & \psi_{2\pi-q}
    \end{pmatrix}
    \begin{pmatrix}
    -h+\cos q & \gamma \sin q\\
    \gamma \sin q & h-\cos q
    \end{pmatrix}
    \begin{pmatrix}
    \psi_q\\ \psi^{\dagger}_{2\pi-q}
    \end{pmatrix}
\end{align}
Here we define
\begin{equation}\label{rotazione Bogoliubov}
    \begin{pmatrix}
    \psi_{q}\\[1ex] \psi^{\dagger}_{2\pi-q}
    \end{pmatrix} \equiv
    \begin{pmatrix}
    \cos\theta_q & \sin \theta_q\\[1ex]
    -\sin \theta_q & \cos\theta_q
    \end{pmatrix}
    \begin{pmatrix}
    \chi_{q}\\[1ex] \chi^{\dagger}_{2\pi-q}
    \end{pmatrix}
\end{equation}
where the angle $\theta_q$ obeys the constraints
\begin{equation}
    \theta_{0,\pi} \equiv 0,
\end{equation}
\begin{equation}
    \begin{cases}
        \sin \theta_{2\pi-q} = -\sin\theta_q\\
        \cos\theta_{2\pi-q} = \cos\theta_q.
    \end{cases}
\end{equation}
The Hamiltonian now reads
\begin{align}
    H^{\pm}  =& \frac{J}{2}\sum_{q\in\Gamma^\pm}\begin{pmatrix}
    \chi_q^{\dagger} & \chi_{2\pi-q}
    \end{pmatrix}
    \Tilde{H}
    \begin{pmatrix}
    \chi_q\\ \chi^{\dagger}_{2\pi-q}
    \end{pmatrix}
\end{align}
with
\begin{equation}
    \Tilde{H} = \begin{pmatrix}
    (-h+\cos q)\cos(2\theta_q)-\gamma\sin q \sin (2\theta_q) & \gamma \cos(2\theta_q)\sin q + (-h+\cos q)\sin(2\theta_q)\\
    \gamma \cos(2\theta_q)\sin q + (-h+\cos q)\sin(2\theta_q) & (h-\cos q)\cos(2\theta_q)+\gamma\sin q \sin (2\theta_q)
    \end{pmatrix}.
\end{equation}
This form brings us to the final step of the diagonalization. One has here to distinguish the ferromagnetic case $J<0$, the most studied on, and the antiferromagnetic case $J>0$, the one that shows frustration.
\subsubsection{Ferromagnetic case}
In the ferromagnetic case ($J<0$), which has been extensively studied \cite{Franchini_2017}, it is convenient to chose $\theta_q$ such that
\begin{equation}\label{def angolo di Bogoliubov}
    e^{i2\theta_q} = \frac{h-\cos q  +i\gamma \sin q}{\sqrt{(h-\cos q)^2 + \gamma^2 \sin^2q}} \qquad q\neq 0,\pi.
\end{equation}
Furthermore, we introduce the dispersion relation
\begin{equation}\label{epsilon(q)}
    \epsilon(q)\equiv \sqrt{(h-\cos q)^2 + \gamma^2 \sin^2q}.
\end{equation}
Note that
\begin{align}
    &\epsilon(\pi) 
    = h+1 \qquad\mbox{if}\quad h\geq 0\\
    &\epsilon(0) 
    = \begin{cases}
    h-1 &\quad\mbox{if}\quad h>1\\
    -h+1&\quad\mbox{if}\quad 0\leq h<1\,.
    \end{cases}
\end{align}
From the choice \ref{def angolo di Bogoliubov} it follows that, independently from the parity of $N$, we have
\begin{align}   H^+=&J\sum_{q\in\Gamma^+}\epsilon(q)\left(\chi_q^\dagger\chi_q-\frac{1}{2}\right)\label{H^+ferro}\\
    H^- =& 
     \begin{cases}
    -J\sum_{q\in \Gamma^-}\epsilon(q)\left(\chi_q^{\dagger} \chi_q-\frac{1}{2}\right) &\quad\mbox{if}\quad  h>1\\[1 ex]
    -J\sum_{q\in \Gamma^-\setminus\{0\}}\epsilon(q)\left(\chi_q^{\dagger} \chi_q-\frac{1}{2}\right) +J\epsilon(0)\left(\chi_0^{\dagger} \chi_0-\frac{1}{2}\right) &\quad\mbox{if}\quad 0\leq h<1 \label{H^-ferro}.
    \end{cases}
\end{align}
\subsubsection{Antiferromagnetic case}
In the antiferromagnetic case ($J>0$) it is convenient to chose $\theta_q$ such that
\begin{equation}\label{def angolo di Bogoliubov antiferro}
    e^{i2\theta_q} = \frac{-h+\cos q  -i\gamma \sin q}{\sqrt{(h-\cos q)^2 + \gamma^2 \sin^2q}} \qquad q\neq 0,\pi.
\end{equation}
For $N$ even one has
\begin{align}
    H^+ =& 
    J\sum_{q\in\Gamma^+}\epsilon(q)\left(\chi_q^\dagger \chi_q-\frac{1}{2}\right)\label{H^+ even}\\
    H^- 
    =& \begin{cases}
    J\sum_{q\in \Gamma^-\setminus\{0,\pi\}}\epsilon(q)\left(\chi_q^{\dagger} \chi_q-\frac{1}{2}\right)-J\epsilon(\pi)\left(\chi_\pi^{\dagger} \chi_\pi-\frac{1}{2}\right)-J\epsilon(0)\left(\chi_0^{\dagger} \chi_0-\frac{1}{2}\right) &\quad\mbox{if}\quad  h>1\\[1 ex]
    J\sum_{q\in \Gamma^-\setminus\{\pi\}}\epsilon(q)\left(\chi_q^{\dagger} \chi_q-\frac{1}{2}\right) -J\epsilon(\pi)\left(\chi_\pi^{\dagger} \chi_\pi-\frac{1}{2}\right) &\quad\mbox{if}\quad 0\leq h<1 .
    \end{cases}\label{H^- even}
\end{align}
For $N$ odd, which is the \textbf{frustrated case} and represents the main focus of this work, one has
\begin{align}
H^+=&J\sum_{q\in\Gamma^+\setminus\{\pi\}}\epsilon(q)\left(\chi_q^\dagger \chi_q-\frac{1}{2}\right)-J\epsilon(\pi)\left(\chi_\pi^{\dagger} \chi_\pi-\frac{1}{2}\right)\label{H^+ odd}\\
    H^-=& \begin{cases}
    J\sum_{q\in \Gamma^-\setminus\{0\}}\epsilon(q)\left(\chi_q^{\dagger} \chi_q-\frac{1}{2}\right)-J\epsilon(0)\left(\chi_0^{\dagger} \chi_0-\frac{1}{2}\right) &\quad\mbox{if}\quad  h>1\\[1 ex]
    J\sum_{q\in \Gamma^-}\epsilon(q)\left(\chi_q^{\dagger} \chi_q-\frac{1}{2}\right)  &\quad\mbox{if}\quad 0\leq h<1 
    \end{cases}\label{H^- odd}.
\end{align}

\subsection{Hallmarks of topological frustration}\label{Hallmarks of topological frustration}
The physical properties of the ground state and of the low energy sector in the ferromagnetic and in the antiferromagnetic case can be rather different.
Indeed, only the odd $N$ case in the antiferromagnetic regime is affected by TF, \textit{i.e.} the geometry of the system does not allow us to simultaneously minimize all local interactions in our Hamiltonian \eqref{xy model}. This incompatibility between local and global order induces an excitation in the ground state, which becomes gapless in the thermodynamic limit ($N\to\infty$) \cite{Dong_2016, Dong_2017}, giving rise to the peculiar physics of this system. In this subsection, based on the exact solution just described, we briefly comment on the features of the frustrated XY chain which are independent from the specific (finite) number of sites. For the non-frustrated cases we also refer to the vast literature \cite{Franchini_2017}.
To start, we address the ground state. We call $\ket{GS'^\pm}$, $\ket{GS^\pm}$  and $\ket{GS}$ the most general ground state elements of, respectively,  $H^\pm$, $\frac{1\pm\Pi^z}{2}H^\pm$ and $H$. We use an analogous notation for the corresponding energy $E$. Furthermore, we denote with $\ket{0^\pm}$ the vacuum of fermions $\chi_q$. The strategy we adopt to obtain the ground state is the following: identify $\ket{GS'^\pm}$; extract $\ket{GS^\pm}$ from the states found in the previous step; find the ground state energy $E=\min\{E^+,E^-\}$ and, as a consequence, $\ket{GS}$. In the thermodynamic limit, this algorithm can be realized in a fully analytical way \cite{Franchini_2017,Catalano22, shaikh2023phase}, but at finite $N$ one has to resort to numerical methods.
\\From Eq. \eqref{H^+ferro} and \eqref{H^-ferro} one can see that in the ferromagnetic chain we have
\begin{align}
    &\ket{GS^+} =\ket{0^+},\label{GS^+ even}\\
    &\ket{GS^-}= \chi^\dagger_0\ket{0^-}.   
    \label{GS^- even}
\end{align}
In the same way, from Eq. \eqref{H^+ even} and \eqref{H^- even} in the non frustrated antiferromagnetic chain we have
\begin{align}
    &\ket{GS^+} =\ket{0^+},\label{GS^+ even}\\
    &\ket{GS^-}= \chi^\dagger_\pi\ket{0^-}.   
    \label{GS^- even}
\end{align}
Hence, without frustration we have always
\begin{align}
    &E^+=  -\frac{1}{2}\sum_{q\in\Gamma^+}\epsilon(q),\label{E^+ even}\\
    &E^- = \begin{cases} -\frac{1}{2}\sum_{q\in\Gamma^-}\epsilon(q) + \epsilon(0) &\quad\mbox{if}\quad h\geq 1\\
    -\frac{1}{2}\sum_{q\in\Gamma^-}\epsilon(q) &\quad\mbox{if}\quad 0\leq h<1.\label{E^- even}
    \end{cases}
\end{align}
In words, when $N$ is even the ferromagnetic and antiferromagnetic cases are completely equivalent, with energies given by 
\eqref{E^+ even} and \eqref{E^- even}. It's important to underline that in the absence of frustration only two states alternate as ground state, with $\lfloor N/2\rfloor$ different crossing lines in the first quarter of the parameter space (as we will see in the next Section). Furthermore one can observe from \eqref{epsilon(q)} that $h^2+\gamma^2=1$ is an exact degeneracy line for all $N$.
To determine precisely which is the ground state once fixed $(h,\gamma)$ one has to compare the energies \eqref{E^+ even} and \eqref{E^- even} \cite{depasquale}.

\begin{figure}[h!]
    \centering
    \includegraphics[scale=0.3]{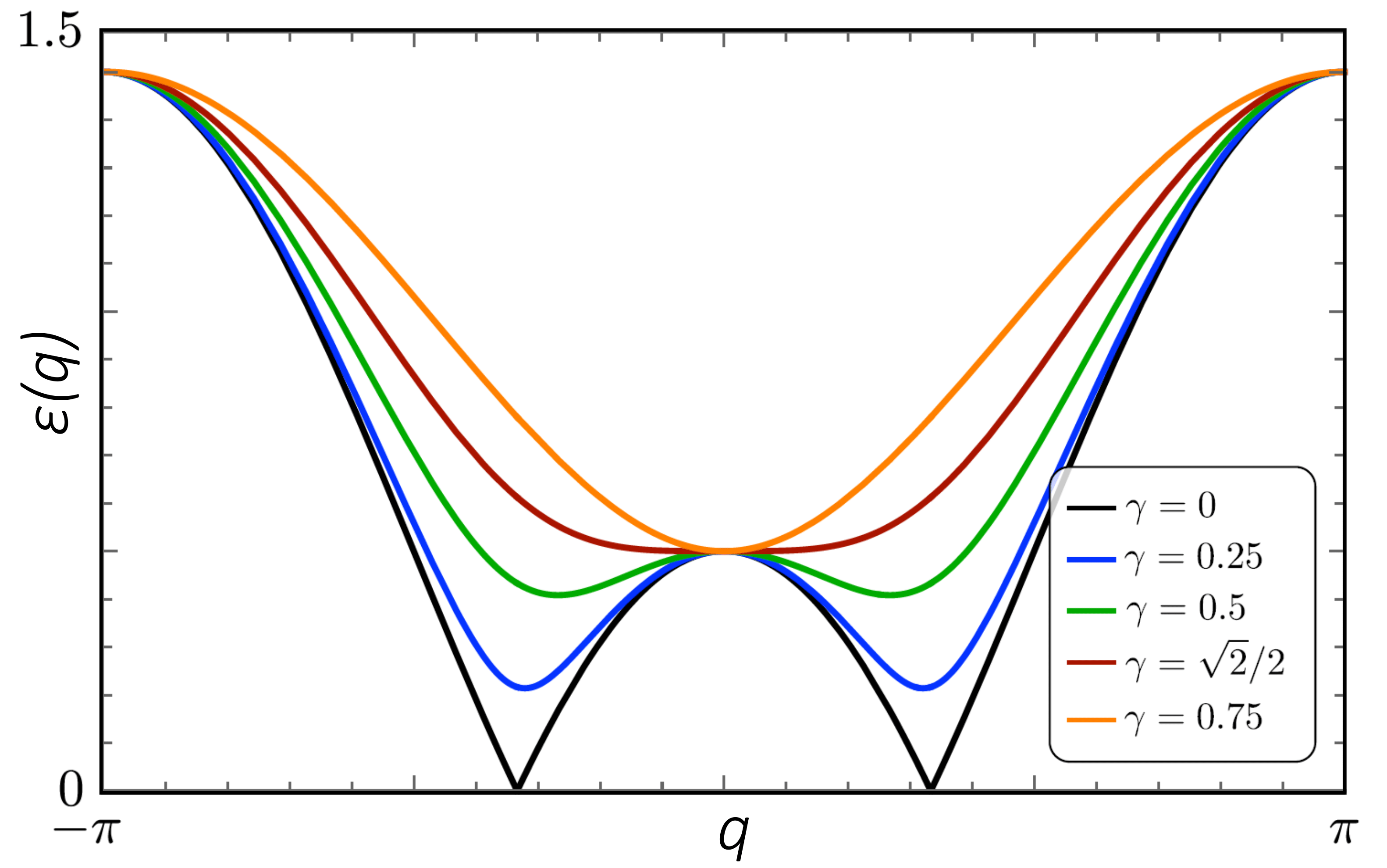}
    \caption{Plot of the dispersion relation $\epsilon(q)$ as a function of $q\in [-\pi,\pi]$ for $h=0.5$ and different values of $\gamma$: $\gamma= 0$ (black), $\gamma=0.25$ (blue), $\gamma=0.5$ (green), $\gamma=\gamma^* (0.5)=\sqrt{2}/2$ (red) and $\gamma=0.9$ (orange).}
    \label{fig:grafico_epsilon_h<1}
\end{figure}

It can be shown that by increasing $N$ the energy gap between these two states closes exponentially \cite{Cabrera1987, Damski_2014}, giving rise to a doubly degenerate manifold which spontaneously breaks ${Z}_2$ symmetry \cite{Franchini_2017}.
Note that TF, that is the impossibility (due to geometry) to minimize simultaneously all local interactions in the Hamiltonian, is translated into the fact that (when $\abs{h}<1$, which is the region of interest) we cannot chose $\ket{GS'^\pm}$ as ground states having them, respectively, $z$-parity equal to $\mp1$. In other words, the lowest energy states of $H^+$ and $H^-$ are not compatible with the $z$-parity constraint \eqref{z parity constraint}. This fact implies that the number of states involved at zero temperature scale as $N$ because of the quantization of momenta (see   Eq. \eqref{def Gamma^pm}). In the doubly degenerate regions of the parameter space the ground space is spanned by two states with non zero equal and opposite momenta \cite{Catalano22, shaikh2023phase}. This is a consequence of the mirror symmetry of our theory, as proved by Mari\'c \textit{et al.} in \cite{QPT_induced_by_TF}.
\\Notice that in absence of external magnetic field $h$ we have $[H,\Pi^\alpha] =0$ (\textit{i.e.} each parity operator is a symmetry), furthermore in the odd $N$ case the parity operators satisfy the non-commuting algebra $\{\Pi^\alpha, \Pi^\beta\} = 2\delta_{\alpha,\beta}$. As a consequence, if we consider a state $\ket{\psi}$ with definite energy and $z$-parity, the $\Pi^x\ket{\psi}$ has the same energy but opposite $z$-parity. Hence the absence of $h$ and the oddity of $N$ imply that even at finite size each state is always degenerate an even number of times (Kramers degeneracy) and we have no need of the thermodynamic limit to get spontaneous symmetry breaking. Taking advantage of this extra symmetry, Mari\'c \textit{et al.} \cite{QPT_induced_by_TF,Frustration_of_being_odd} managed to compute analytically the magnetization, defined by $m_j^\alpha \equiv \expval{\sigma_j^\alpha}$, in the frustrated case. For $\gamma>1$, where the ground space is two times degenerate, they found that the average magnetization is constant and is suppressed by the total system size as $\frac{1}{N}$. They call it \textit{mesoscopic ferromagnetic magnetization}. Instead, on the $(h,\gamma)=(0,\gamma<1)$ line the ground state is four times degenerate and, as previously observed, the operators $H$, $\Pi^z$ and the translation operator $T$ (defined by $T^\dagger \sigma_j^\alpha T \equiv \sigma_{j+1}^\alpha$) form a set of compatible operators. As a consequence, one has two possibilities: either fixing a translational invariant state and having the mesoscopic ferromagnetic magnetization or giving up translational invariance and flowing into a state whose magnetization looks like the staggered one but changes incommensurably over the chain \cite{Frustration_of_being_odd,QPT_induced_by_TF,fate_of_local_order}. On the contrary, in the absence of frustration the magnetization at zero external magnetic field is zero in the even $N$ cases and ferromagnetic for the even $N$ ferromagnetic chain.
\\By increasing the number of sites in the frustrated case, we expect that there exists a critical value $\gamma^*(N,h)$ such that when $0<\gamma< \gamma^*(N,h)$ the number of regions in the parameter space alternating in $z$-parity would increase \cite{Catalano22}. 
Such an expectation is justified by the fact that when $0<\gamma<\gamma^*$ (and $0<h<1$) the dispersion relation \eqref{epsilon(q)} has the shape in Figure \ref{fig:grafico_epsilon_h<1}, where the minima are at $p(h,\gamma) = \pm\arccos{\frac{h}{1-\gamma^2}}$ \cite{Dong_2017}. By increasing $N$, the cardinality of $\Gamma^+$ and $\Gamma^-$ increases linearly, as a consequence  by moving in this subregion of the parameter space we have a more and more frequent change of the elements of $\Gamma^+$ and $\Gamma^-$ which are closest to $p(h,\gamma)$ \cite{Catalano22, shaikh2023phase}. 


To conclude this introductory Section, observe that at the point $(h,\gamma)=(0,1)$ the Hamiltonian \eqref{xy model} reduces to that of the classical Ising chain. As consequence, the ground space in the frustrated case is $2N$-fold degenerate and spanned by
the “kink states”, which have a single pair of nearest neighbor
spins (the kink) that are ferromagnetically aligned in the $x$ direction and the other $N - 1$ pairs of
nearest-neighbor spins antiferromagnetically aligned. Without frustration, the ground space at the classical point is always two times degenerate and spanned by the two Neél states in the antiferromagnetic case and by the two completely $x$-ferromagnetic states in the ferromagnetic case. The signature of the different value of the degeneracy can be found in the Taylor expansions around the classical point in the two cases.

\section{Finite size effects}\label{Finite size ground state comparison}
In this Section we compare the finite size low energy states of the frustrated case with that of the unfrustrated counterparts. Although we only focus on few particular values of $N$, the analysis can be straightforwardly extended to general values of the number of sites \cite{Catalano22}.

\subsection[$N=6$]{\boldmath$N=6$}
\begin{figure}[h]
    \centering
    \includegraphics[scale=0.55]{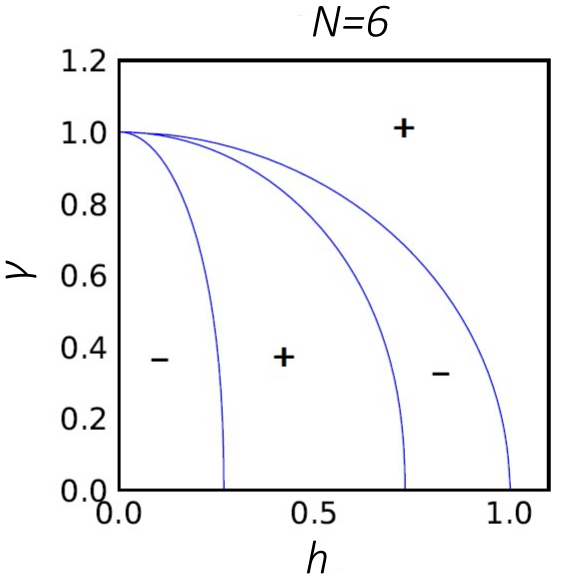}
    \caption{Double degeneracy curves (in blue) in $N=6$ case. This plot is independent from the sign of $J$. The signs $+$ and $-$ specify the $z$-parity of the (unique) ground state in that region.}
    \label{contour N=6}
\end{figure}

\begin{figure}[h]
    \centering
    \includegraphics[scale=0.4]{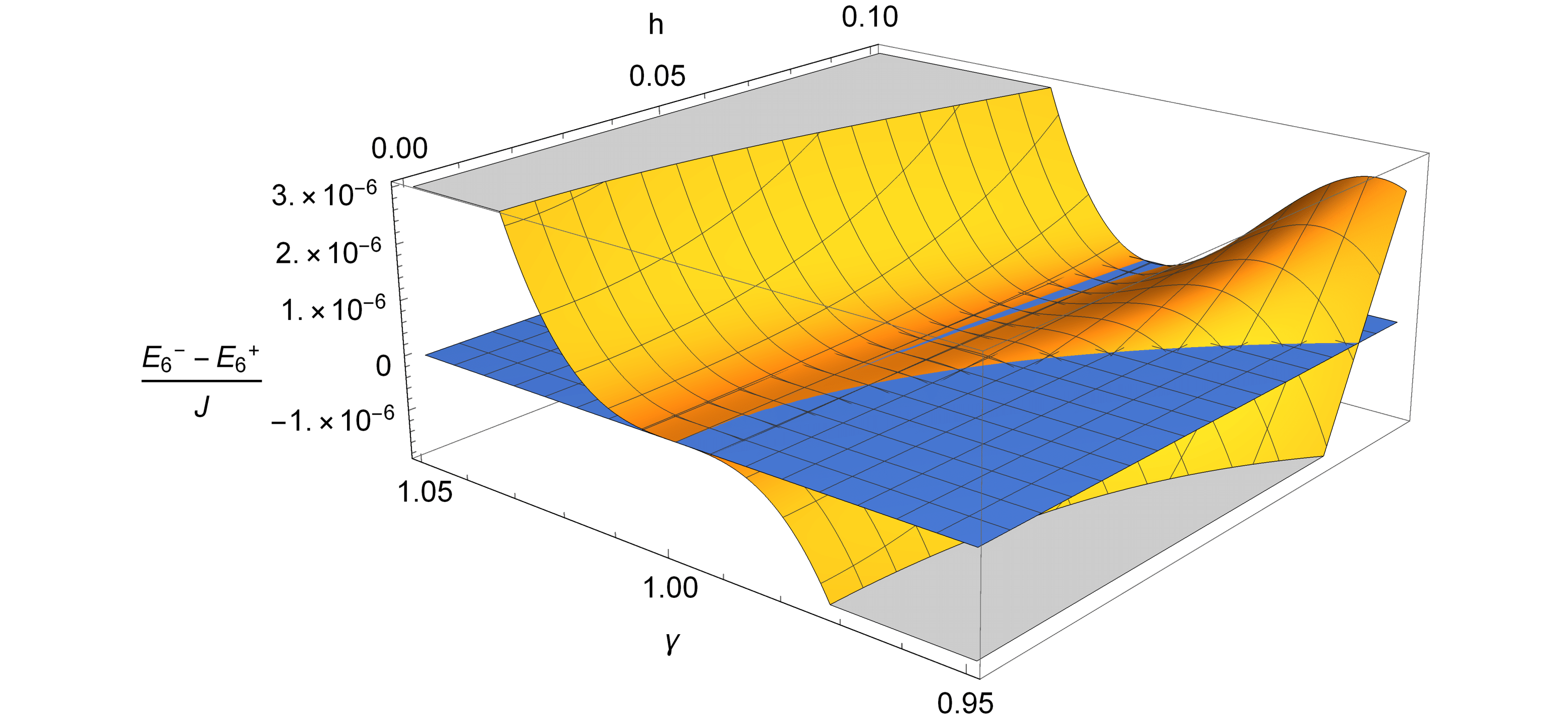}
    \caption{Difference between $E_6^-$ and $E_6^+$ at sixth order in Taylor series around the classical point (in orange) compared with the plane $E_6^- - E_6^+=0$ (in blue).}
    \label{Taylor N=6}
\end{figure}

The results, which (as already underlined in the previous Section) do not depend on the sign of $J$, are reported in Figure \ref{contour N=6}, where we have drawn in blue the curves of exact double degeneracy
of the ground-state (where $GS = \mbox{span}\{\ket{GS^+}, \ket{GS^-}\}$)
and we specified the $z$-parity of the ground state vector in the non-degeneracy regions.
\\From Figure \ref{contour N=6}  we can see that the neighbourhood of the classical point $(h,\gamma) = (0,1)$ (where the model reduces to the classical Ising ring) with $h>0$ is divided in four regions with alternated $z$-parities (remember that when $\Pi^z=+1$ or $\Pi^z=-1$ the ground state is given, respectively, by \eqref{GS^+ even} or \eqref{GS^- even}), so we expect that the  first order Taylor expansions of $E^+$ and $E^-$ around the classical point are equal. Our expectations are confirmed by numerical results, telling us that
\begin{align}\label{eq Taylor even}
    E_6^- - E_6^+ = \frac{3}{32}(\gamma-1)^3 + \order{(\gamma-1)^4, h^2(\gamma-1)^2},
\end{align}
whose plot up to sixth order in the parameters is in figure \ref{Taylor N=6}.

 We also observed that
\begin{equation}
    E_6^-(h,\gamma=1) - E_6^+(h,\gamma=1) \approx \frac{63}{256}h^6.
\end{equation}

\subsection[$N=5$ ferro]{\boldmath$N=5$ ferro}
\begin{figure}[h]
    \centering
    \includegraphics[scale=0.53]{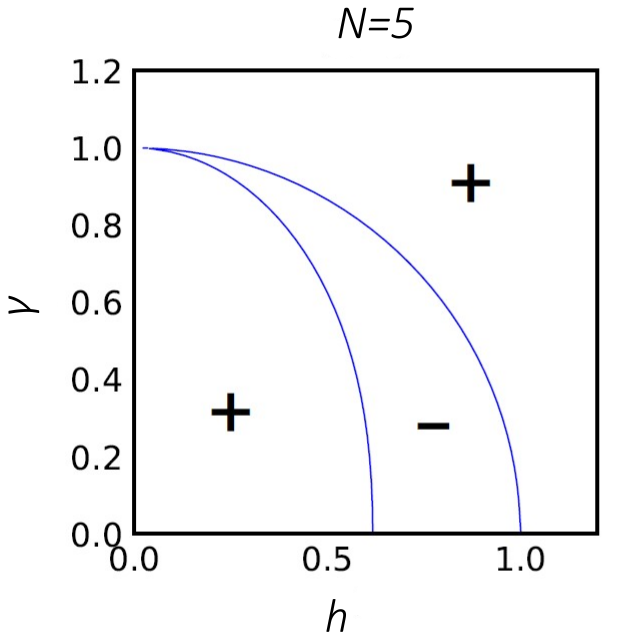}
    \caption{Double degeneracy curves (in blue) in $N=5$ ferromagnetic case for $h\neq 0$. The signs $+$ and $-$ specify the $z$-parity of the (unique) ground state in that region.}
    \label{contour N=5 ferro}
\end{figure}

\begin{figure}[h]
    \centering
    \includegraphics[scale=0.4]{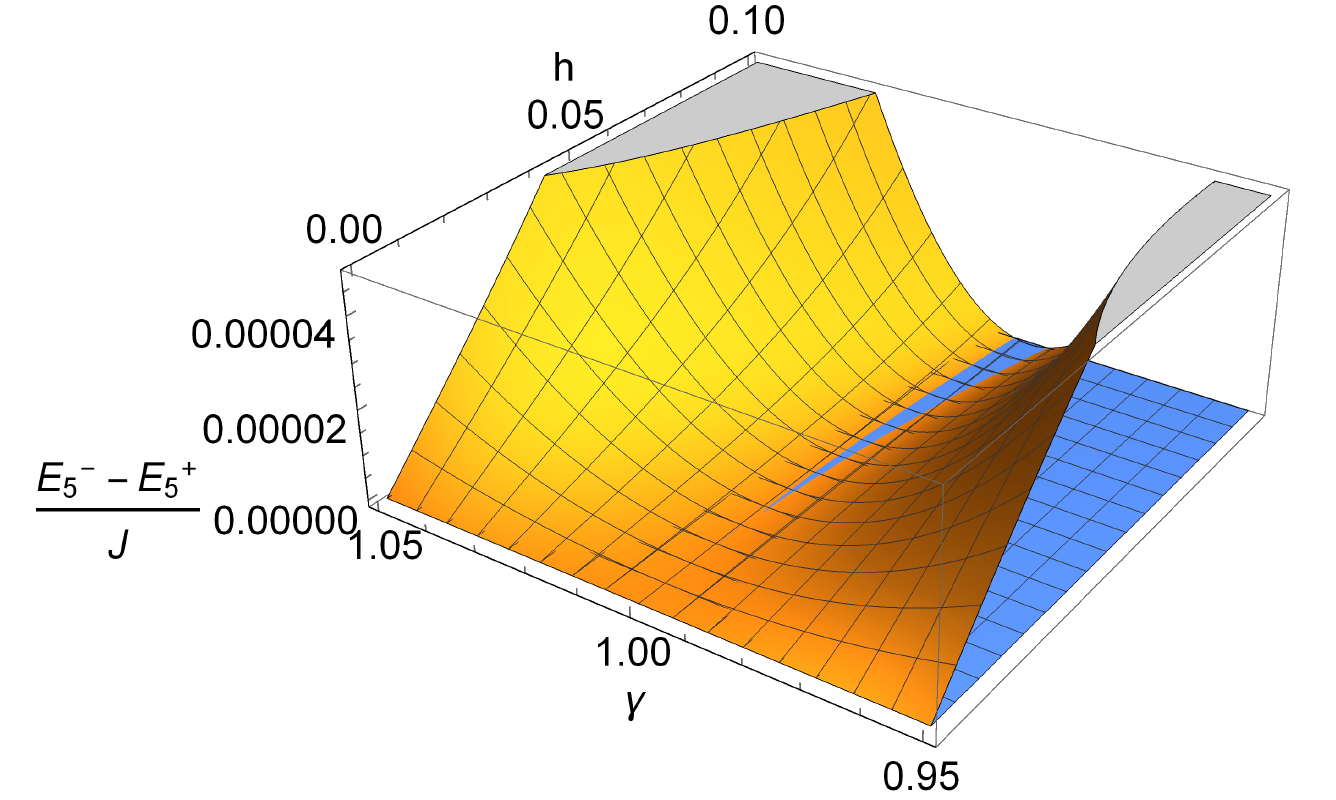}
    \caption{Difference between $E_5^-$ and $E_5^+$ at sixth order in Taylor series around the classical point (in orange) compared with the plane $E_5^- - E_5^+=0$ (in blue).}
    \label{Taylor Ferro 5}
\end{figure}

The results are reported in Figure \ref{contour N=5 ferro}, where we have drawn in blue the curves of exact double degeneracy of the ground-state (where $GS = \mbox{span}\{\ket{GS^+}, \ket{GS^-}\}$)
and we specified the $z$-parity of the ground state vector in the non-degeneracy regions.
From Figure \ref{contour N=5 ferro}  we can see that the neighbourhood of the classical point $(h,\gamma) = (0,1)$ with $h>0$ is divided in three regions with alternated $z$-parities, so we expect that the  first order Taylor expansions of $E^+$ and $E^-$ around the classical point are equal (exactly as in the $N=6$ case). Our expectations are confirmed by numerical results, telling us that
\begin{equation}
    E_5^- - E_5^+ = \frac{15}{32}h(\gamma-1)^2 + \order{h(\gamma-1)^3, h^3(\gamma-1)},
\end{equation}
whose sixth order plot is in Figure \ref{Taylor Ferro 5}, which is similar to Figure \ref{Taylor N=6} except for the fact that, as already stressed in Section \ref{Hallmarks of topological frustration}, the oddness of $N$ implies the exact double degeneracy at $h=0$. 

We also observed that
\begin{equation}
    E_5^-(h,\gamma=1) - E_5^+(h,\gamma=1) \approx \frac{35}{128}h^5.
\end{equation}

\subsection[$N=5$ antiferro]{\boldmath$N=5$ antiferro}
\begin{figure}[h]
    \centering
    \includegraphics[scale=0.55]{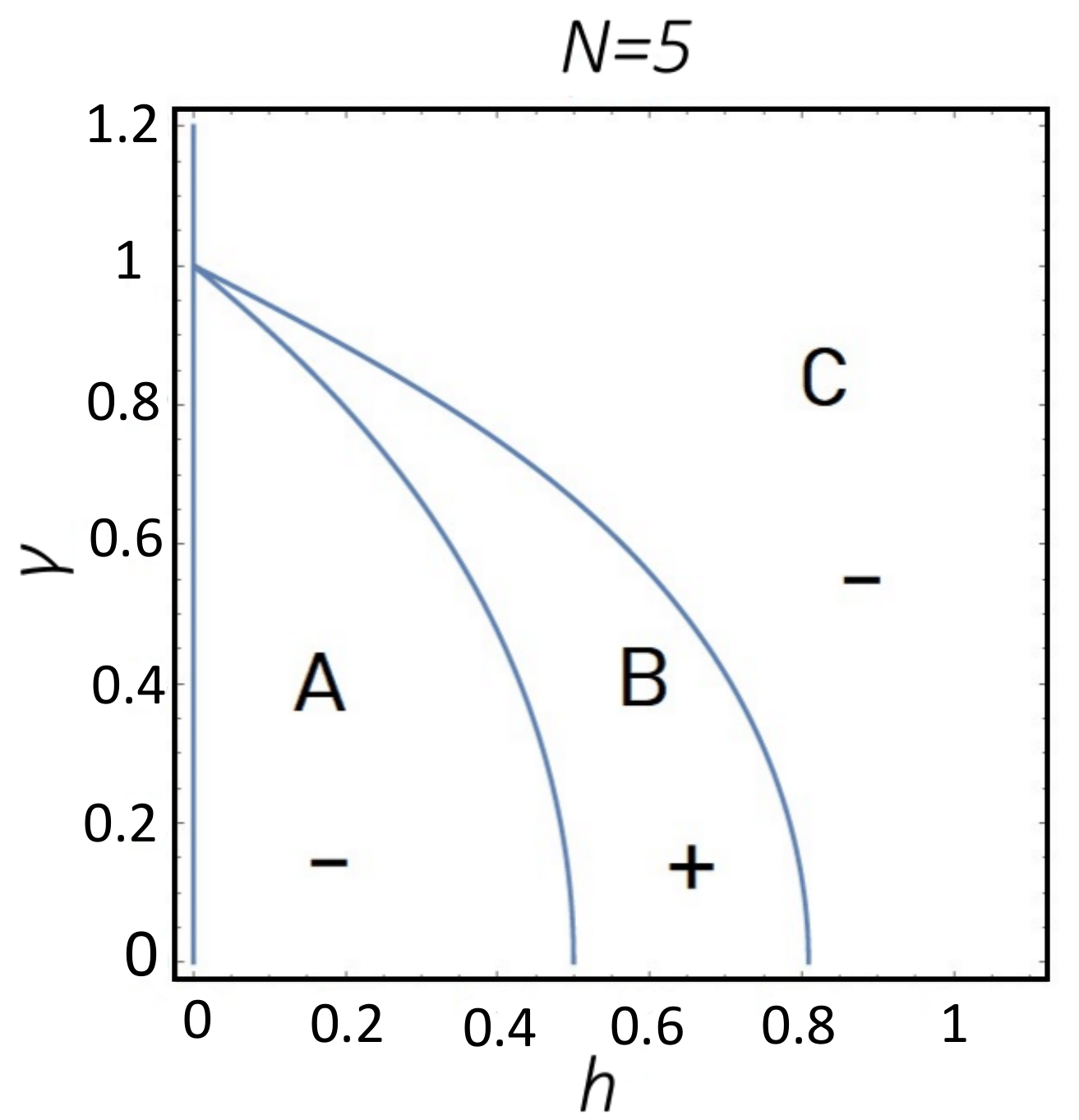}
    \caption{Degeneracy lines (in blue) in $N=5$ case. The letters $A$, $B$ and $C$ label the three regions with three different ground state vectors in which the first quarter of the parameter space is divided. As in the even $N$ case, the signs $+$ and $-$ specify the $z$-parity of the (unique) ground state in that region.}
    \label{contour N=5}
\end{figure}

\begin{figure}[h!]
    \centering
    \includegraphics[scale=0.32]{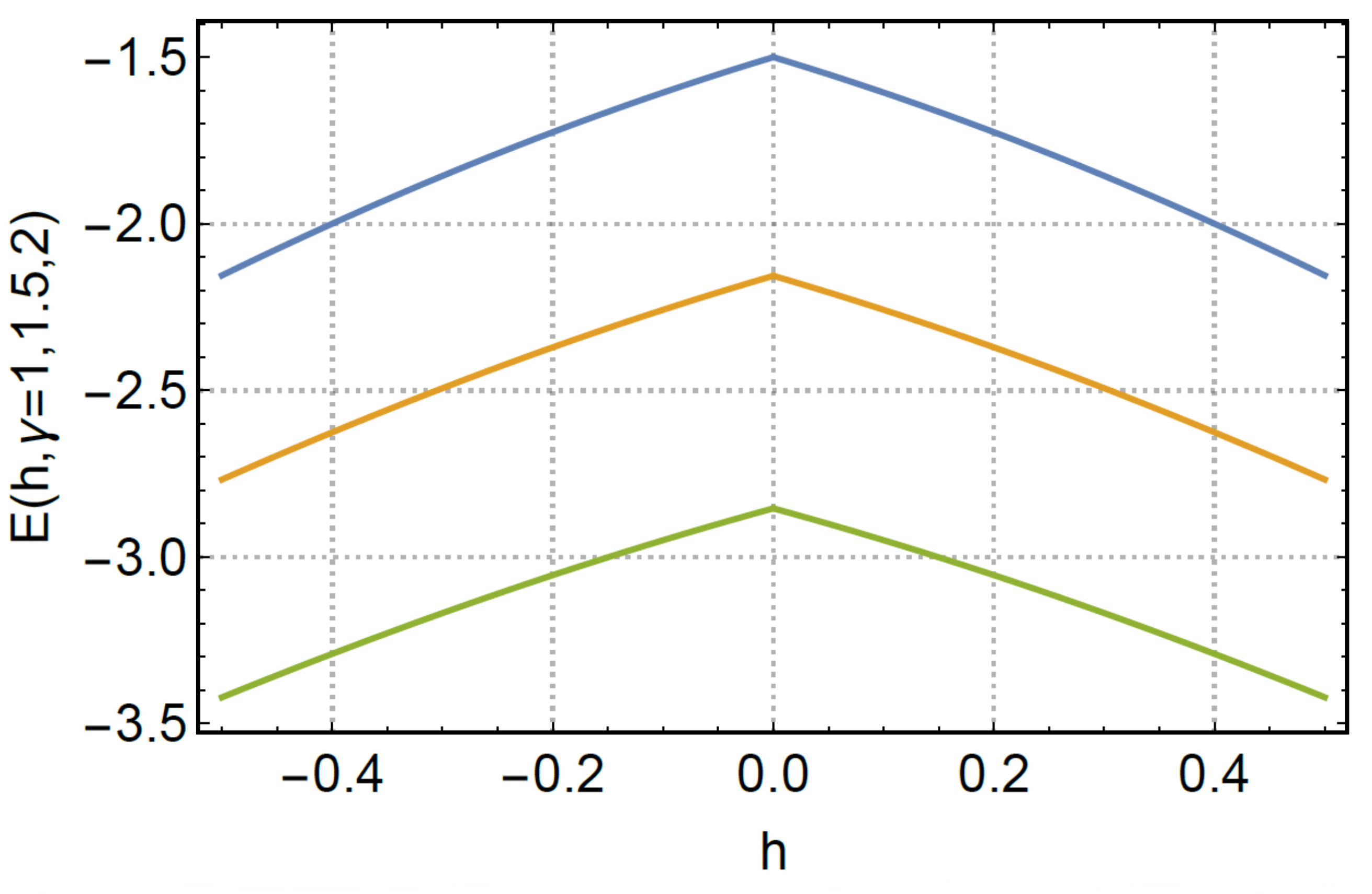}
    \caption{Ground state energy $E$ in $N=5$ case as a function of $h\in(-0.5,0.5)$ for $\gamma$ equal to 1 (blue), 1.5 (orange) and 2 (green).}
    \label{E odd discontinuity}
\end{figure}

\begin{figure}[h!]
\centering
\includegraphics[scale=0.27]{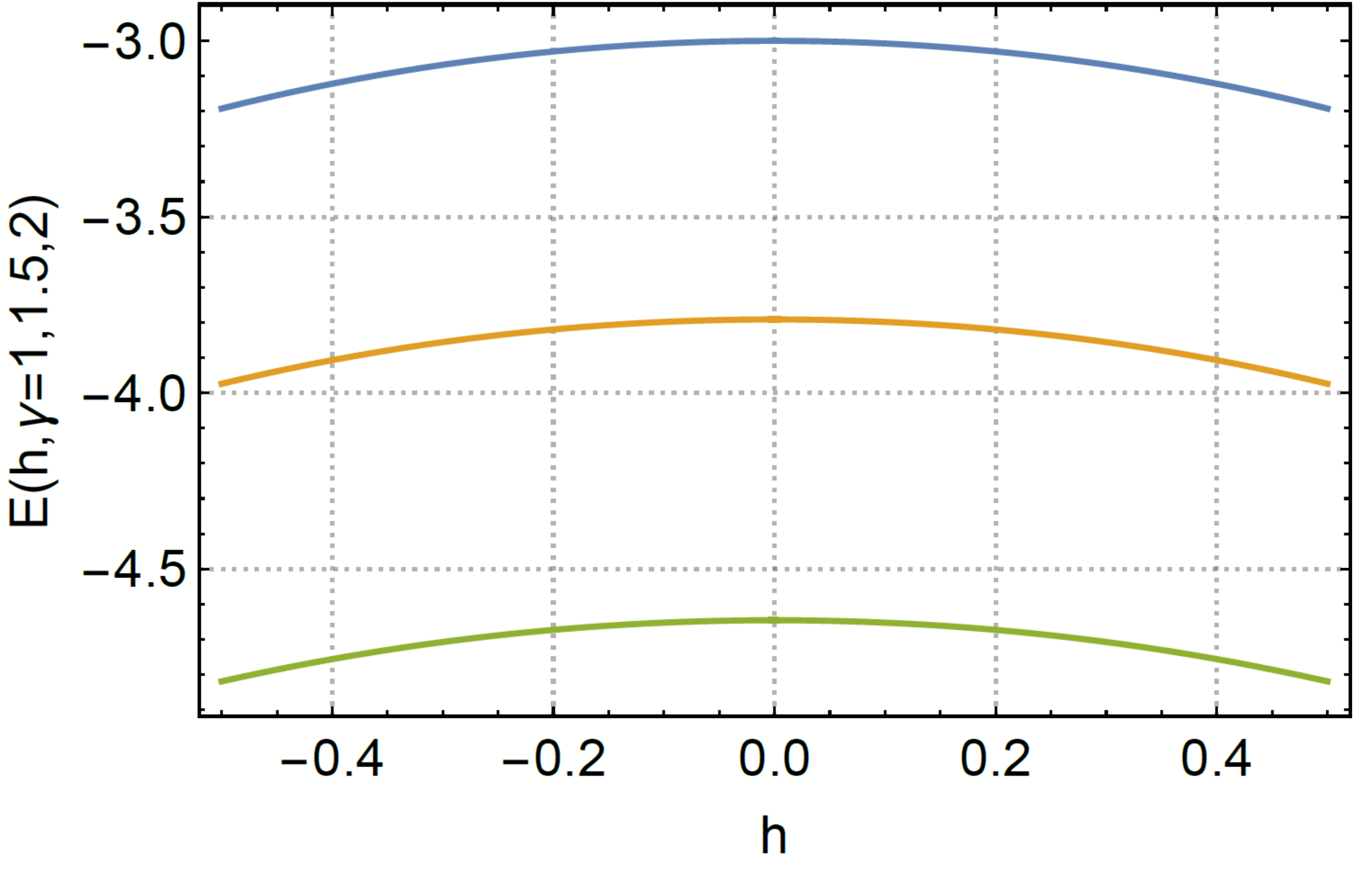}
\caption{Ground state energy $E$ in the absence of frustration as a function of $h\in(-0.5,0.5)$ for $\gamma$ equal to 1 (blue), 1.5 (orange) and 2 (green). Moreover, $N=6$. The picture applies to both the ferromagnetic and the antiferromagnetic case}
\label{fig:E even smooth}
\end{figure}

\begin{figure}[h!]
\centering
\includegraphics[scale=0.27]{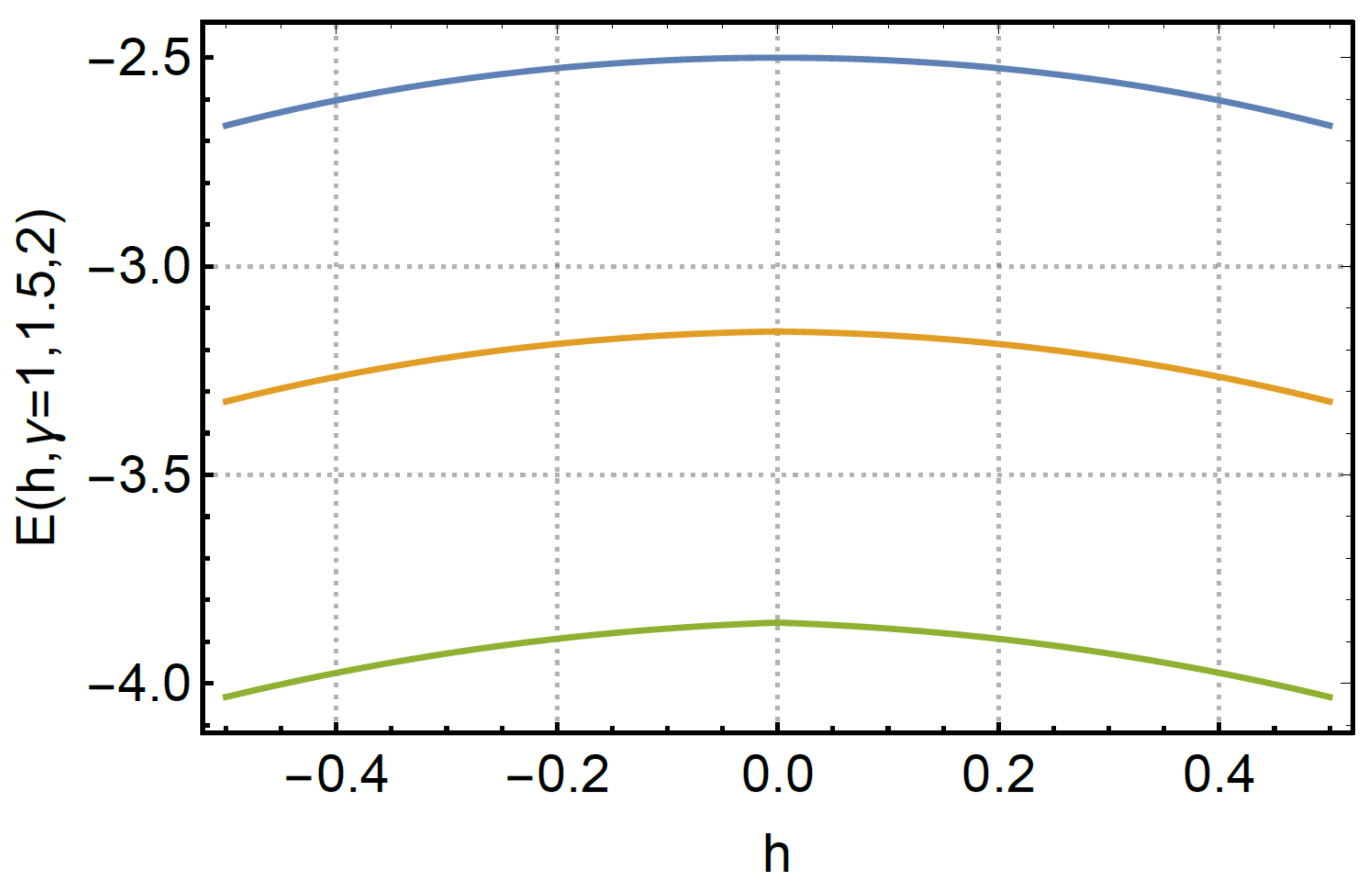}
\caption{Ground state energy $E$ in the absence of frustration as a function of $h\in(-0.5,0.5)$ for $\gamma$ equal to 1 (blue), 1.5 (orange) and 2 (green). Moreover, $N=5$ and the coupling is ferrmoagnetic.}
\label{fig:E odd ferro smooth}
\end{figure}

We find numerically that when $N=5$ the ground state is
\begin{align}
    GS = 
    \begin{cases}
        \mbox{span}\big\{ \chi_0^{\dagger}\ket{0^-}, \ket{0^+}\big\} &\qquad h=0,\,\gamma>1\\[1.6 ex]
        \mbox{span}\big\{ \chi_{2\pi/5}^{\dagger}\ket{0^-}, \chi_{-2\pi/5}^{\dagger}\ket{0^-},\chi_{3\pi/5}^{\dagger}\chi^\dagger_\pi\ket{0^+},\chi_{-3\pi/5}^{\dagger}\chi^\dagger_\pi\ket{0^+}\big\} &\qquad h=0,\,0<\gamma<1\\[1.6 ex]
        \mbox{span}\big\{ \chi_{2\pi/5}^{\dagger}\ket{0^-}, \chi_{-2\pi/5}^{\dagger}\ket{0^-}\big\} &\qquad (h,\gamma)\in A\\[1.6ex]
   \mbox{span}\big\{ \chi_{\pi/5}^{\dagger}\chi^\dagger_\pi\ket{0^+}, \chi_{-\pi/5}^{\dagger}\chi^\dagger_\pi\ket{0^+}\big\} &\qquad(h,\gamma)\in B\\[1.6ex]
\chi_0^{\dagger}\ket{0^-} &\qquad (h,\gamma)\in C, \label{gs odd}
    \end{cases}
\end{align}
where the regions $A$, $B$ and $C$ refer to Figure \ref{contour N=5}. 
\\The ground state energy is
\begin{align}
    E = 
    \begin{cases}
        -\frac{1}{2}\sum_{q\in\Gamma^-}\epsilon(q)+\epsilon\left(\frac{2}{5}\pi\right)  \qquad &h=0,\,0<\gamma<1 \cup (h,\gamma)\in A\\[1.6 ex]
        -\frac{1}{2}\sum_{q\in\Gamma^+}\epsilon(q)+\epsilon\left(\frac{\pi}{5}\right)    \qquad &(h,\gamma)\in B\\[1.6 ex]
        -\frac{1}{2}\sum_{q\in\Gamma^-}\epsilon(q)+ \theta(1-h)\epsilon(0)  \qquad &h=0,\,\gamma>1 \cup (h,\gamma)\in C
        \label{energy gs odd}
    \end{cases}
\end{align}
where $\theta(x)$ is the Heaviside step function. It's easy to observe that \eqref{energy gs odd} is non analytical at $(h,\gamma)=(0,\gamma\geq 1)$, having a discontinuity in its first derivative with respect to $h$ (see Figure \ref{E odd discontinuity}), which survives  in the thermodynamic limit, giving birth to a first order boundary quantum phase transition \cite{articolo_tesi}. On the contrary, this doesn't happen in the even $N$ case and in the odd ferromagnetic case, where the ground state energy \eqref{E^+ even} is always analytic at $(h,\gamma)=(0,\gamma\geq 1)$, as shown, respectively, in Figures \ref{fig:E even smooth} and \ref{fig:E odd ferro smooth}. 


For $0<\gamma<1$ and  $\gamma>1$ the ground state's degeneracy is, respectively, 4 and 2 \cite{QPT_induced_by_TF,Frustration_of_being_odd}. Notice also that at the boundary between regions $A$ and $B$ and regions $B$ and $C$ the ground state is degenerate, respectively, four and three times. 

\begin{figure}[H]
    \centering
\includegraphics[scale=0.5]{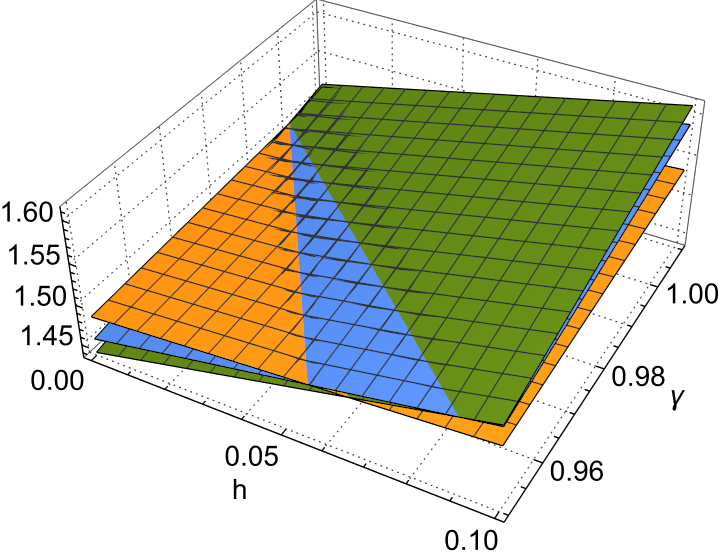}
    \caption{Plot of Taylor expansion of $-E_A$ (orange), $-E_B$ (blue) and $-E_C$ (green) truncated to first order in $h$ and $\gamma-1$ as a function of $h$ and $\gamma$ around the classical point.} 
    \label{Taylor dispari}
\end{figure}

In the neighbourhood of the classical point $(h,\gamma)=(0,1)$ with $h>0$ the different alternating quantum states in the $N=5$ case are three (and given, explicitly, by \eqref{gs odd}). The energies of these states which, for the sake of simplicity, we will call $\ket{A}$, $\ket{B}$ and $\ket{C}$ (with reference to the subregions of Figure \eqref{contour N=5}) have the following first order Taylor expansions around the classical point
\begin{eqnarray}
    E_A &=& -\frac{3}{2}+\frac{1-\sqrt{5}}{4}h +\frac{5-\sqrt{5}}{8}(\gamma-1) + \order{h^2,(\gamma-1)^2, h(\gamma-1)}\label{E_A}\\
    E_B &=& -\frac{3}{2}-\frac{1+\sqrt{5}}{4}h -\frac{5+\sqrt{5}}{8}(\gamma-1) + \order{h^2,(\gamma-1)^2, h(\gamma-1)}\label{E_B}\\
    E_C &=& -\frac{3}{2}-h -\frac{5}{4}(\gamma-1) + \order{h^2,(\gamma-1)^2, h(\gamma-1)}\label{E_C},
\end{eqnarray}
where $h>0$. The situation is depicted in Figure \ref{Taylor dispari} where, to simplify the visualization, we reported the first order of the Taylor expansions of $-E_A$, $-E_B$ and $-E_C$.

From \eqref{E_A}, \eqref{E_B} and \eqref{E_C} we observe that
\begin{equation}
    E_i-E_j = \alpha_{ij}h + \beta_{ij}(\gamma-1) + \order{h^2,(\gamma-1)^2, h(\gamma-1)} \qquad i,j=A,B,C\quad i\neq j\,,
\end{equation}
where $\alpha_{ij}$ and $\beta_{ij}$ are different from zero.

\section{Implementation on the quantum computer}

This section addresses the experimental investigation of the results found by Dong \textit{et al.} in \cite{Dong_2016}. In particular, we briefly introduce the technique used to evaluate the physical observables of interests and discuss the outcomes from both numerical simulations and runs on a real IBM quantum device.

\subsection{Variational Quantum Eigensolver}
The Variational Quantum Eigensolver (VQE) \cite{peruzzo} is a hybrid quantum–classical variational algorithm that produces an upper-bound estimate of the ground-state $|\Psi_{GS}\rangle$ and its energy $E_{GS}$ of a Hamiltonian $H$ \cite{westerheim2023dualvqe}. 
Based on Rayleigh-Ritz variational principle, it has found ubiquitous application in fields ranging from quantum chemistry \cite{magnetochemistry} to nuclear physics \cite{nucphysvqe} and many-body physics \cite{lmg_grossi}. The algorithm is actually hybrid, involving both quantum and classical operations. More specifically, VQE proceeds according to the following workflow:
\begin{enumerate}
    \item An ansatz wavefunction $|\Psi \left( \pmb{\theta}  \right) \rangle$ is prepared on a quantum computer through a parametric quantum circuit; 
    \item The energy for the given ansatz is evaluated $\displaystyle E_{|\Psi \left(\pmb{\theta} \right) \rangle} = \frac{\langle \Psi \left( \pmb{\theta} \right)|H|\Psi\left( \pmb{\theta} \right) \rangle}{\langle \Psi \left( \pmb{\theta} \right)|\Psi \left( \pmb{\theta} \right) \rangle} \geq E_{GS}$;
    \item The set of parameters specifying the state, here generically indicated with $\pmb{\theta}$, is updated in order to minimize the energy through a classical optimization routine.
\end{enumerate}
The choice of the ansatz is crucial for the whole algorithm to work properly. The employed $|\Psi \left( \pmb{\theta} \right) \rangle$ must indeed meet a reasonable tradeoff between expressivity and trainability: a great number of parameters certainly allows to represent more diverse wavefunctions (including the optimal one), but can significantly slow down the minimization process. 
Being a Parametrized Quantum Circuit (PQC), it could be characterized by barren plateaus where the gradient becomes exponentially small in the number of qubits during the training. Here, not only the choice of the ans\"atze is important but also the initialization strategy of its parameters deserves a theoretical discussion for practical applications \cite{initialization}.

There is a huge amount of literature for specific use cases, here we cite two approaches to potential design strategy: physically motivated ones \cite{physmot1} \cite{physmot2}, which are usually difficult to implement on near term quantum devices, and hardware efficient ans\"atze (HEA) \cite{hea} that are devised to limit the consequences of noise.\\
As for the present work, a heuristic trial wavefunction consisting of layered $SU(2)$ and $CX$ gates has been employed and the optimization procedure has been carried out through a stochastic gradient descent (SGD) method named SPSA \cite{SPSA}.

\subsection{Ideal simulations and results}
\begin{figure}[H]
    \centering
\includegraphics[scale=0.5]{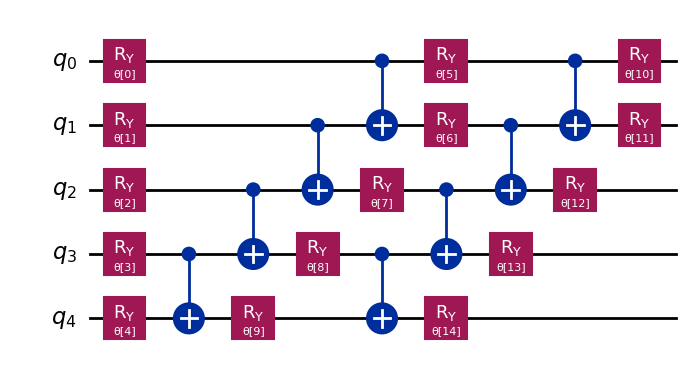}
    \caption{Example of ansatz for the VQE algorithm. The angles of rotations are left as free parameters to be optimized during the procedure and linear entanglement has been chosen, with $CX$ gates linking adjacent qubits.} 
    \label{ansatz}
\end{figure}

We start our investigation by leveraging ideal simulations of the finite-size model, noiseless setups serving as a perfect benchmark to explore the possible extension of the study to real device runs. For this reason, we consider the Hamiltonian
\begin{align}
    H = J \sum^N_{i=1} \left[ \sigma_i^x\sigma_{i+1}^{x} + h \sigma^z_i \right]
\end{align}
for $N=5,6$ qubits and periodic boundary conditions. We consider the $J=1$ and $J=-1$ cases separately and evaluate the ground state energy for $h \in \lbrace -0.3, 0, 0.3\rbrace$. 
The VQE procedure is started from a two-layers ansatz as the one showed in figure \ref{ansatz} with random initial parameters$\pmb{}$. 

\begin{figure}[H]
    \centering
\includegraphics[scale=0.45]{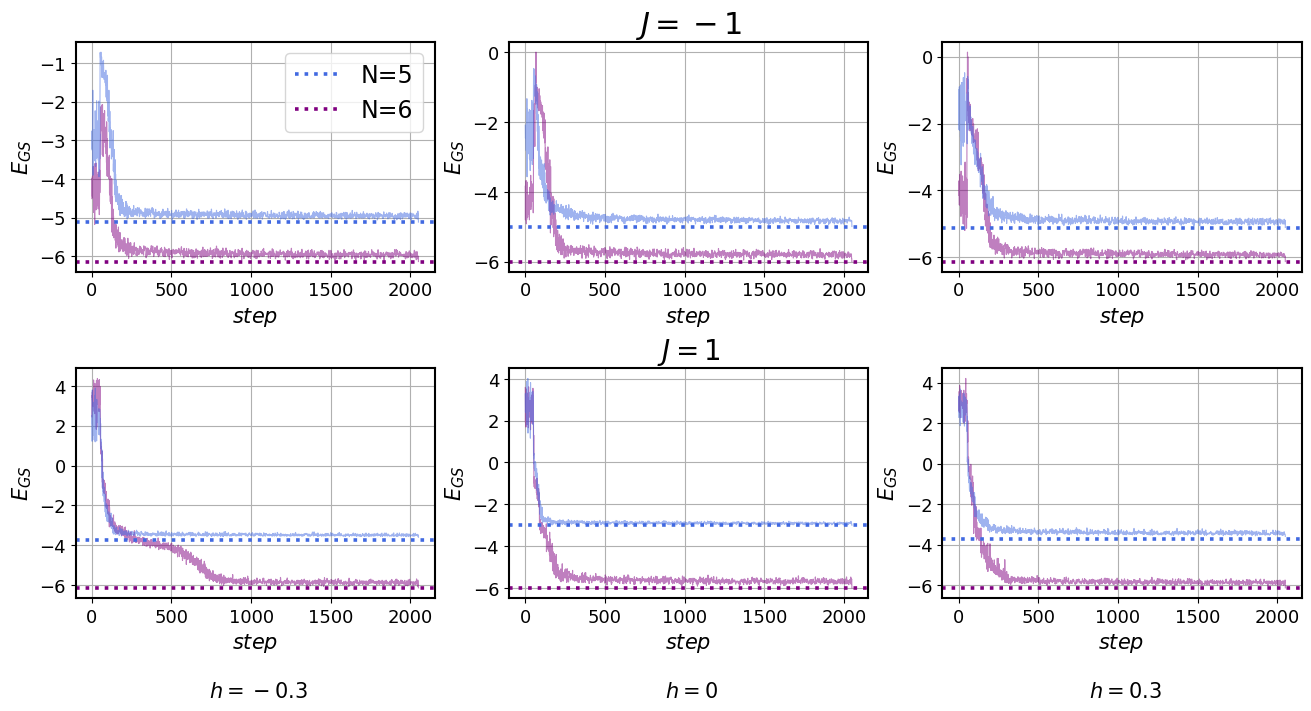}
    \caption{Estimated $E_{GS}$ during a $2000-$steps optimization procedure with SPSA for both $J=-1,1$ and $N=5,6$. Dotted lines refer to the exact ground state energies. } 
    \label{idealconv}
\end{figure}

The approximate values of $E_{GS}$ at each minimization step for every choice of $J, N$ and $h$ are reported in figure \ref{idealconv}, ensuring a good and fast convergence with respect to the exact values in dotted line.

\begin{figure}[h!]
    \centering
\includegraphics[scale=0.45]{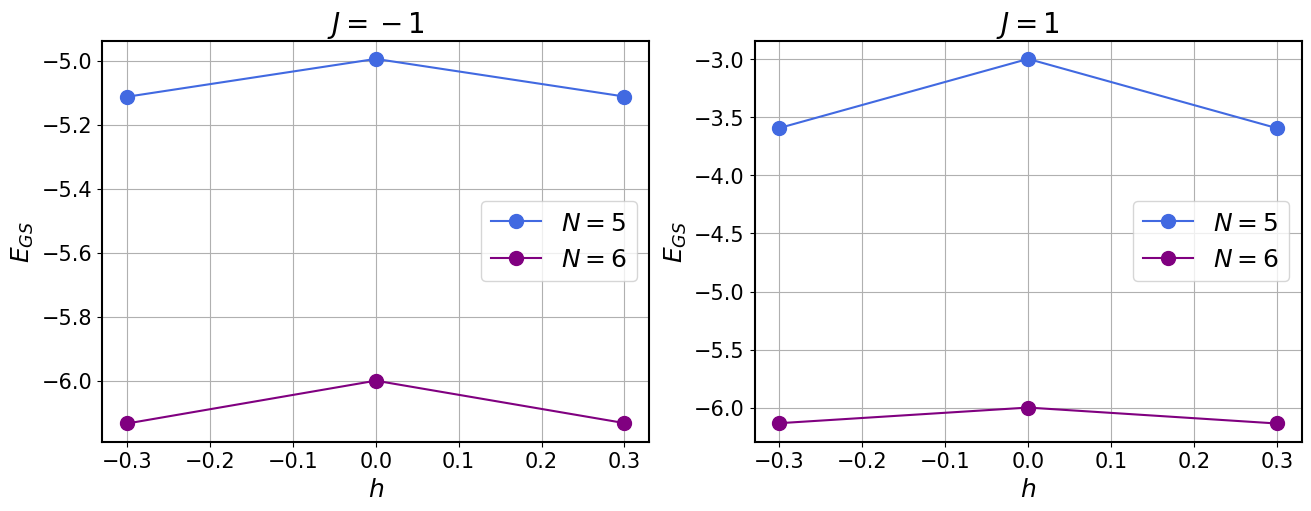}
    \caption{$E_{GS}$ obtained through VQE as a function of the magnetic field $h$ for each realization of $J$ and $N$. } 
    \label{idealen}
\end{figure}

In light of evaluating the derivative difference $\displaystyle\frac{\partial E_{GS}}{\partial h}\big|_{h=0^+} - \displaystyle\frac{\partial E_{GS}}{\partial h}\big|_{h=0^-}$, we consider the output  $E_{GS}\left( h\right)$ of the VQE algorithm as displayed in figure \ref{idealen} for a small deviation of the magnetic field from the null value.

Using forward difference to compute numerical derivatives (table \ref{ideal_der_table}), we find satisfactory agreement with analytical predictions as regards both numerical values and the role of $J$ in determining the behaviour of $E_{GS}$ for even and odd values of $N$. 

\begin{table}[H]
\centering
\resizebox{5cm}{!}{
\begin{tabular}{ |c |c| c|  }
\hline 
  & $N=5$  & $N=6$ \\
  \hline
 $J=1 $ &  $-3.96$ & $ -0.904$  \\
 \hline
 $J=-1$ &    $-0.786$ &  $-0.888$ \\
 \hline\hline
\end{tabular}

}
\caption{\label{ideal_der_table}{Derivative differences obtained from noiseless simulations.
}}
\end{table}

\subsection{Superconducting quantum device results}
\begin{figure}[h!]
    \centering
\includegraphics[scale=0.3]{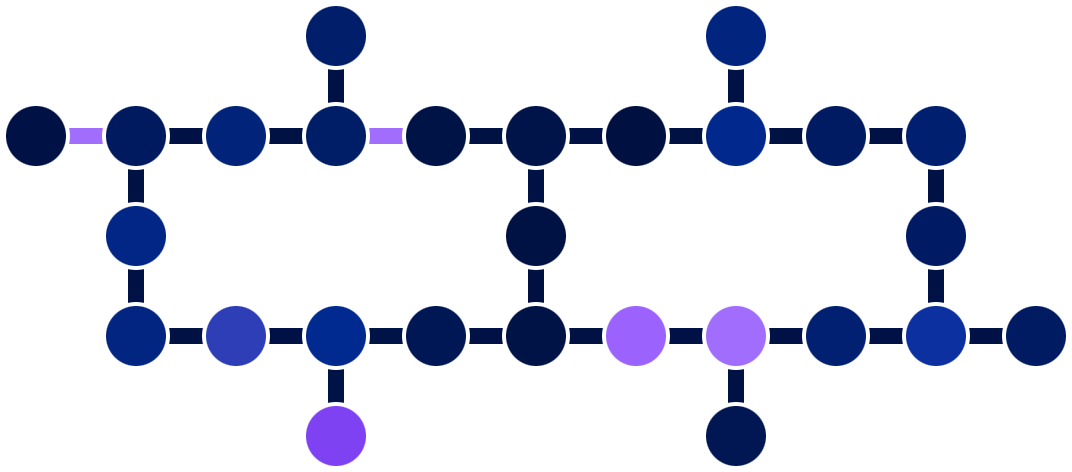}
    \caption{Schematic representation of the topology of the $27-$qubits \textit{ibm-cairo} device, on which computations were performed. Calibration data, at the time of running were: Median T1: $89.55 us$, Median T2: $98.7 us$, Median CNOT error: $9.668e-3$ and Median readout error: $1.470e-2$. } 
    \label{ibmcairo_top}
\end{figure}

In the following we show and discuss results obtained from a superconducting transmon IBM Quantum chip. Starting from encouraging preliminary results, in terms of performance and evaluation of $E_{GS}$ in the previously described noiseless scenario, we extend our analysis further by analysing stability and reproducibility in a real context, where a real, noisy quantum computer is used, and where state of art error suppression techniques are exploited.
The quantum device used in this work, namely \textit{ibm-cairo} (Falcon $r5.11$) 
consists of $27$ fixed-frequency transmons qubits, with fundamental transition frequencies of approximately $5$ GHz and anharmonicities of $-340$ MHz. Microwave pulses are used for single-qubit gates and cross-resonance interaction for two-qubit gates. The experiments took place without intermediate calibration using IBM Qiskit Primitive where the quantum platform computes the expectation values of the observable (the energy in this case) with respect to the states prepared by the PQC. The topology of the deployed device is displayed in figure \ref{ibmcairo_top}.

\begin{figure}[h!]
    \centering
\includegraphics[scale=0.45]{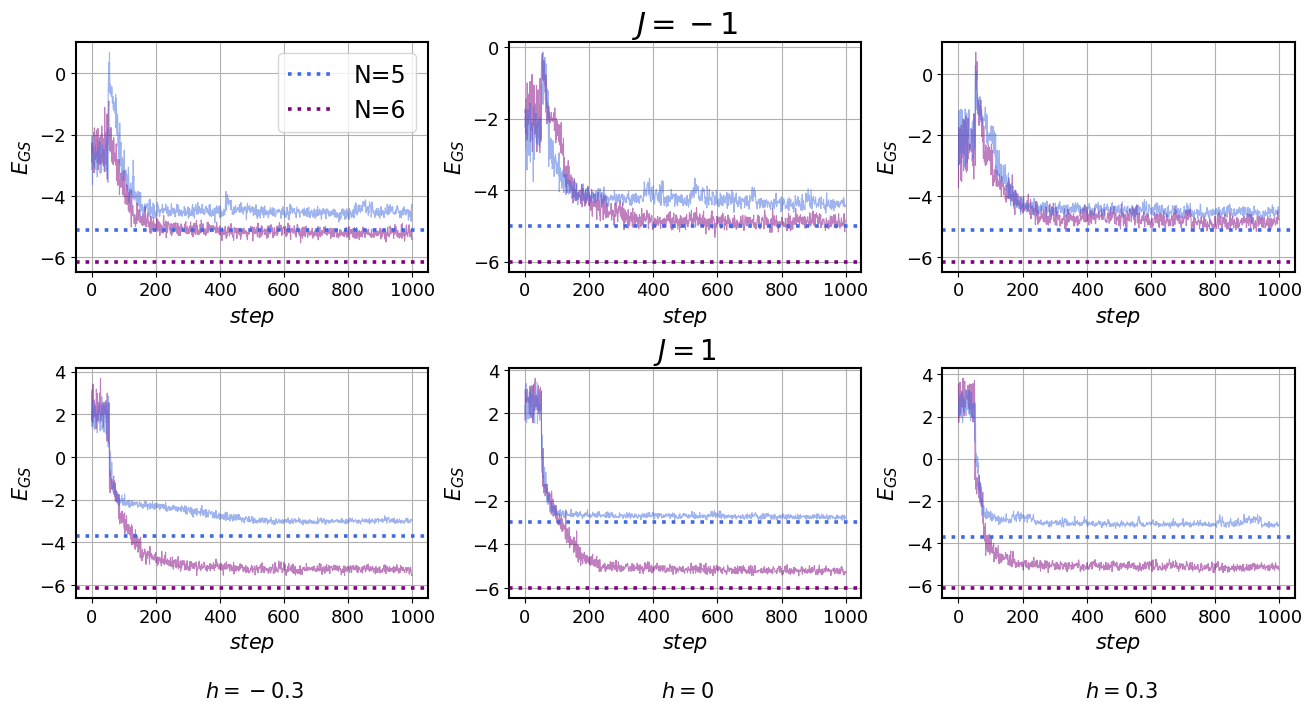}
    \caption{Estimated $E_{GS}$ in a $1000-$steps optimization process for both $J=-1,1$ and $N=5,6$. Dashed lines refer to the exact ground state energies. The number of steps of each optimization procedure is varied in each case.} 
    \label{realhwconv}
\end{figure}

In  case of computation on a quantum hardware it is essential to consider the adverse effects of noise and devise methods to reduce them. For this reason the previous VQE routine is supplemented with error mitigation techniques (EM). In particular, we employ Twisted Readout Error eXtinction (TREX) \cite{trex} and Dynamical Decoupling (DD) \cite{ezzell2023dynamical}. TREX is a model-free approach to readout errors, that result in biases in quantum expectation values as the ones we are interested in. By randomizing the output channel of a circuit through the application of random bit flips prior to measurements, the estimation bias is turned into a multiplicative factor that can be readily divided out.
In addition, DD exploits properly tuned control pulses to average environment induced decoherence to zero, thus contributing to clean out the final result.
Similarly to the noiseless case, the approximate values of $E_{GS}$ as function of the minimization steps for each choice of $J,N$ and $h$ are shown in figure \ref{realhwconv}. 

\begin{figure}[h!]
    \centering
\includegraphics[scale=0.45]{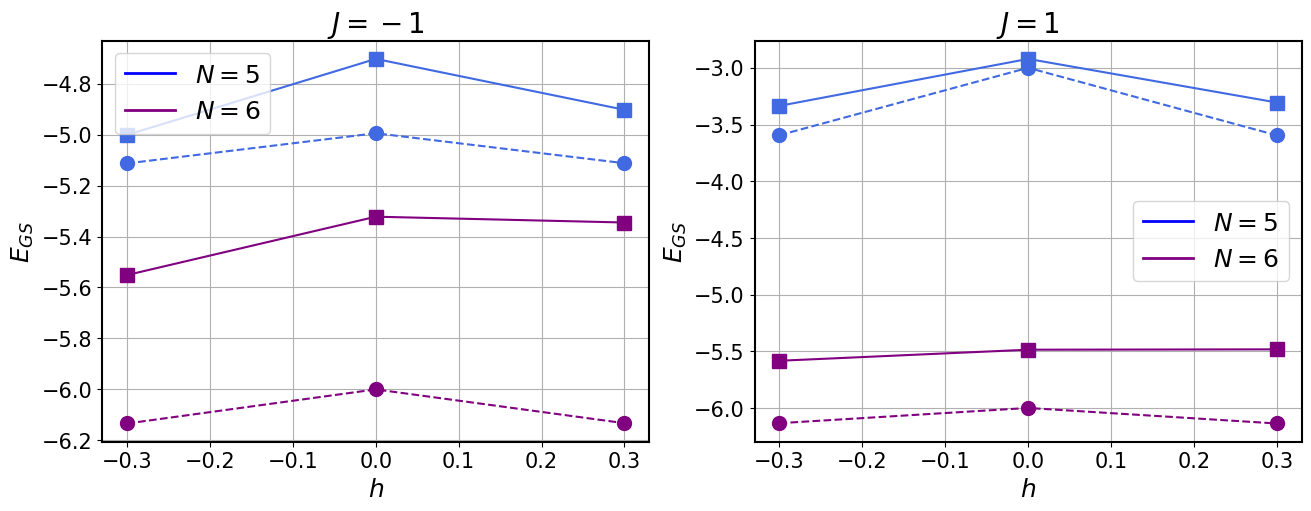}
    \caption{$E_{GS}$ obtained through VQE as a function of the magnetic field $h$ for each realization of $J$ and $N$. Circles and squares respectively  refer to noiseless simulations and real hardware runs. } 
    \label{realhwen}
\end{figure}

As one could expect from the very beginning, the optimization procedure on the quantum platform yields less favorable outcomes than the ideal case, with convergence being achieved at values slightly distant from the exact ones. 
Nonetheless, repeating the protocol used for the ideal simulations, we get what is showed in figure \ref{realhwen}, where the noiseless results are displayed for completeness.

In analogy to the ideal case, the derivative difference is evaluated using the finite difference method.

\begin{table}[H]
\centering
\resizebox{5cm}{!}{
\begin{tabular}{ ||c |c| c|  }
\hline 
  & $N=5$  & $N=6$ \\
  \hline
 $J=1 $ & $-2.66$  & $-0.314$ \\
 \hline
 $J=-1$ & $-1.65$  & $-0.838$  \\
 \hline\hline
\end{tabular}

}
\caption{\label{rhw_der_table}{Derivative differences obtained from runs on the real IBM hardware.
}}
\end{table}

Although decoherence affects the VQE algorithm leading to evaluations of $E_{GS}$ which significantly differs from the exact ones especially for $N=6$ qubits (figure \ref{realhwen}), the results reported in table \ref{rhw_der_table} align satisfactorily with the ones obtained in the noiseless setup, allowing to capture the emergence of the quantum phase transition even on the real hardware.


\section{Conclusion}\label{sec13}
After a detailed presentation of the solution of the XY model with and without frustrated boundary conditions, we compared the low energy states in the two cases resuming some the most relevant results in the literature so far. 
We have then commented in detail the $N=5$ and $N=6$ cases both in the antiferromagnetic and ferromagnetic cases.
Finally we have studied experimentally the Ising chain, implementing the hamiltonian fir different values of the magnetic field with and without frustration for $N=5,6$ on a IBM Quantum computer and our results confirm the theoretical ones found by Dong \textit{et al.} in \cite{Dong_2016}. The experimental results are comforting and confirm the possibility of actually studying physical systems of interest on quantum technologies.

\section*{Authors Contributions}
A.F.M, and F.R.dF equally contributed to the experimental results (Section 4), D.S.S. wrote the theoretical introduction (Sections 2 and 3), N.T.Z. and M.G. equally contributed to the idealization. All Authors took part in the interpretation of the results and in writing the manuscript.
\section*{Acknowledgments}
N.T.Z. acknowledges the funding through the NextGenerationEu Curiosity Driven Project “Understanding even-odd criticality”. N.T.Z. and M.S. acknowledge the
funding through the “Non-reciprocal supercurrent and topological transitions in hybrid Nb-
InSb nanoflags” project (Prot. 2022PH852L) in the framework of PRIN 2022 initiative of the
Italian Ministry of University (MUR) for the National Research Program (PNR). MG is supported by CERN through CERN Quantum Technology Initiative. Access to IBM devices was obtained through the CERN Quantum HUB at CERN.

\section*{Declarations}

The authors declare no competing interests.

\section*{References}
\bibliographystyle{iopart-num}

\providecommand{\newblock}{}


\end{document}